\documentclass[aip,apl,reprint,a4paper]{revtex4-1}
\usepackage{graphicx}                             
\usepackage{epstopdf}
\usepackage{amsmath}
\usepackage{amssymb}

\begin{document}
\title{Spatially resolved investigation of strain and composition variations in (In,Ga)N/GaN epilayers}

\author{Benjamin\ Wilsch}
\author{Uwe\ Jahn}\email{ujahn@pdi-berlin.de}
\author{Bernd\ Jenichen}
\author{Jonas\ L\"ahnemann}
\author{Holger\ T.\ Grahn}
\affiliation{Paul-Drude-Institut f\"ur Festk\"orperelektronik,
Hausvogteiplatz 5--7, 10117 Berlin, Germany}

\author{Hui\ Wang}
\author{Hui\ Yang}
\affiliation{Suzhou Institute of Nanotech and Nanobionics,
Chinese Academy of Sciences, Suzhou 215123, China}

\begin{abstract}
The strain state and composition of a 400~nm thick (In,Ga)N layer grown by metal-organic chemical vapor deposition on a GaN template are investigated by spatially integrated x-ray diffraction and cathodoluminescence (CL) spectroscopy as well as by spatially resolved CL and energy dispersive x-ray analysis.
The CL investigations confirm a process of strain relaxation accompanied by an increasing indium content toward the surface of the (In,Ga)N layer, which is known as the compositional pulling effect. Moreover, we identify the strained bottom, unstrained top, and gradually relaxed intermediate region of the (In,Ga)N layer. In addition to an increase of the indium content along the growth direction, the strain relaxation leads to an enhancement of the lateral variations of the indium distribution toward the surface.
\end{abstract}


\maketitle

The (In,Ga)N material system is a promising candidate for the application in solar cells, since the optical absorption range of this system covers the entire solar spectrum. For this, layers as thick as several 100~nm with significant indium content $x$  ($>$0.1) are necessary. However, the growth of In-rich layers has proven to be difficult, \cite{morales2009} which is primarily due to the lack of suitable substrates as well as the large lattice mismatch between (In,Ga)N and commonly used substrates such as GaN templates.
Previous investigations of such thick layers, e.g., by Wang $et$ $al$.~\cite{wang1,wang2} have shown that strain relaxation by the formation of dislocations occurs above a critical layer thickness. This causes a significant degradation of the structural, optical, and electrical properties of the (In,Ga)N film. Such a variation in the strain state ultimately leads to a variation in the composition of the epilayers. These variations have been observed as two distinct peaks corresponding to different strain states in a reciprocal space map  recorded by x-ray diffraction (XRD) and as two distinct peaks in the optical emission spectra of the thick (In,Ga)N film.~\cite{pereira1,pereira2,pereira3,pereira4,wang1,wang2} The correlation of strain and composition variations of thick (In,Ga)N layers has been reported by a number of groups and has been named by Hiramatsu $et$ $al$.~\cite{hiramatsu1997} the "compositional pulling-effect''.  This effect describes a reduction of the indium incorporation during the early stages of the (In,Ga)N layer growth on lattice-mismatched substrates such as GaN. As the growth proceeds, the (In,Ga)N layer begins to relax, subsequently allowing for higher indium incorporation rates.~\cite{hiramatsu1997} However, XRD and spatially integrated luminescence experiments average laterally and vertically over large regions. Therefore, they do not allow for a clear determination of whether the regions of varying strain and/or indium composition are distributed vertically or laterally. Thus, complementary spatially resolved investigations such as cathodoluminescence (CL) spectroscopy in a scanning electron microscope (SEM) are desirable in order to confirm the conclusions drawn from the XRD results.

We present spatially resolved investigations of the compositional pulling effect for a 400~nm thick (In,Ga)N layer grown on a GaN template by metal-organic chemical vapor deposition (MOCVD). To this end, we use CL spectroscopy and x-ray microanalysis in an SEM to determine the vertical and lateral variations of the indium content of the (In,Ga)N layer. Our results confirm previous conclusions about the indium and strain distribution along the growth direction drawn from XRD investigations of (In,Ga)N layers with varying thickness.~\cite{pereira1,pereira2,wang1} Moreover, we show that during growth of thick (In,Ga)N layers the strain relaxation leads not only to a distinct distribution of the indium content, but also to a significant decrease of the quantum efficiency of the layer within a depth range, where the strain relaxation is expected to occur. Furthermore, we see an increase of lateral variations of the indium content for the relaxed upper part of the layer.

The sample was grown in a horizontal MOCVD reactor using c-plane sapphire as the substrate. In a first step, a $2$~$\mu$m GaN layer was grown on sapphire to serve as a pseudo-substrate for the (In,Ga)N growth. Before the nominally 400~nm thick (In,Ga)N layer was deposited, the as-grown GaN surface was thermally annealed at $700$~$^{\circ}$C in an NH$_3$ atmosphere. The precursors were triethylgallium (TEGa), trimethylindium (TMIn), and ammonia.
To allow for a detailed analysis of the variation in strain and composition across the (In,Ga)N layer, cross-sections of the sample were prepared using an ion beam in a Gatan Ilion$^{+}$ device. Structural images were recorded with a ZEISS Ultra-55 SEM. The SEM is further equipped with a Gatan MonoCL3 system consisting of a slide-in parabolic mirror and an optical spectrometer to collect and disperse the CL signal, respectively. Subsequently, the CL signal is detected either by a fast photomultiplier for monochromatic CL imaging or by a liquid nitrogen-cooled charge-coupled device array to acquire CL spectra. The x-ray microanalysis has been performed utilizing an EDAX Apollo XV detector with a super ultra-thin window (S-UTW) for energy dispersed x-ray (EDX) measurements. For CL and EDX, the applied electron beam energy amounted to 5 and 7~keV, respectively. These values represent a compromise between a satisfactory signal-to-noise ratio, which requires a sufficiently high beam energy, and a high spatial resolution, for which the beam energy should be small. The spatial resolution for the chosen beam energy was estimated using CASINO,~\cite{hovington,demers} a Monte-Carlo simulation software for electron scattering within the sample. For the CL experiment, we must also account for the diffusion of the excited electron-hole pairs, which results in a resolution of about 50~nm for both CL and EDX measurements.

\begin{figure}
\centerline{\includegraphics*[width=7cm]{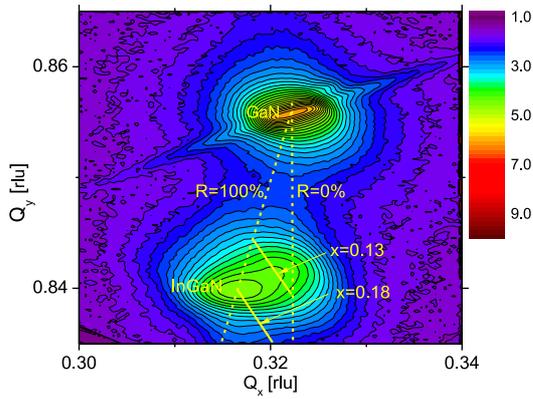}}
\caption{(Color online) Reciprocal space map of the (In,Ga)N/GaN layer structure. $R$ and $x$ represent the degree of relaxation and indium content, respectively.}
\label{fig1}
\end{figure}

The spatially averaged investigation of the strain and indium content has been performed by analyzing a reciprocal space map acquired by XRD using  the Cu$_{\mbox{\scriptsize K}_{\alpha_1}}$ radiation ($\lambda=0.154$~nm) of a PANalytical X'Pert Pro System equipped with a Ge ($220$) hybrid monochromator.
In Fig.~\ref{fig1}, the reciprocal space map of the sample is shown. In addition to the reflection of the GaN buffer, it exhibits a double peak structure for the (In,Ga)N layer. Analysing this double peak structure according to our previous work~\cite{wang1,wang2} and following the procedure described by Pereira $et$ $al$.,~\cite{pereira1,pereira2,pereira3} we conclude that the first peak corresponds to an indium content of $x=0.13$ and a degree of relaxation of $R=25$\%, while the second peak corresponds to $x=0.18$ and $R= 100$\%. This is consistent with the known relation that an advanced relaxation is accompanied by an increase of the indium incorporation.
Since the strain relaxation sets in after a critical thickness has been exceeded during growth, it is usually assumed that the strain variation and thus the variation of the indium content occurs along the growth direction. In order to confirm this assumption, we performed spatially resolved CL and EDX investigations.

\begin{figure}[!b]
\centerline{\includegraphics*[width=6cm]{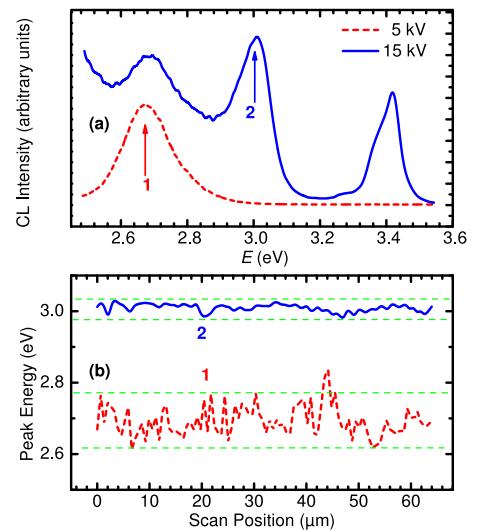}}
\caption{(Color online) (a) CL spectra of the (In,Ga)N/GaN structure excited from the top using an electron beam energy of 5 (dashed line) and 15~keV (solid line). (b) Peak energy of the CL lines marked by ``1'' (dashed line) and ``2'' (solid line) in (a) as a function of the scan position of the electron beam along a line on the surface of the (In,Ga)N layer. All data have been acquired at 300~K.}
\label{fig2}
\end{figure}

A rough assessment of vertical strain- and/or indium content-related energy variations can be obtained by CL measurements from the top of the sample for different electron beam energies. Figure~\ref{fig2}(a) shows two CL spectra, where a 20$\times$20~$\mu$m$^2$ large region of the sample was excited from the top and the excitation depth was varied by using beam energies of 5 and 15~keV. For 15~keV, the CL spectrum consists of a GaN-related line centered at 3.42~eV and two (In,Ga)N-related peaks centered at 2.68 and 3.01~eV labeled ``1'' and ``2''  in Fig.~\ref{fig2}(a).  The increase of the CL signal for energies below 2.5~eV is caused by the GaN-related yellow luminescence. Thus for 15~keV, the excitation volume covers the whole (In,Ga)N layer and even penetrates into the GaN buffer layer. Excitation using 5~keV results in a spectrum, which consists only of the (In,Ga)N-related peak centered at 2.68~eV. From this result, we can already conclude that the part of the (In,Ga)N layer close to the surface emits light at lower energies than the part close to the GaN/(In,Ga)N interface, which can be due to both an increase of the indium content and a strain relaxation toward the surface.
In order to obtain information about the lateral energy variations in the (In,Ga)N layer, we have measured CL spectra along a 70~$\mu$m long line on the surface of the (In,Ga)N layer again using both beam energies 5 and 15~keV. Figure~\ref{fig2}(b) shows the energies of the two CL peaks as a function of the scan position. We conclude that the lateral energy variations are considerably smaller than the ones in the vertical direction. However, the lateral energy variations close to the surface are about 3 times larger than the ones found in the bottom part of this layer.

\begin{figure}
\centerline{\includegraphics[width=9cm]{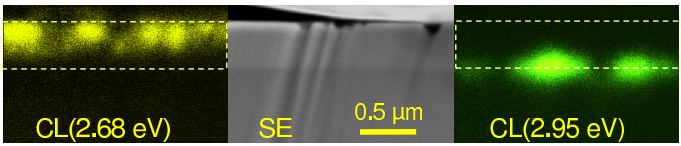}}
\caption{(Color online) Secondary electron (center) and CL (left and right) images of the cross-section of the (In,Ga)N/GaN structure acquired at 300~K. The CL images differ in the used CL detection energy, which lies within the peaks ``1'' and ``2'' of Fig.~\ref{fig2}(a).}
\label{fig3}
\end{figure}

 Figure~\ref{fig3} shows a secondary electron (center) and two CL (left and right) images of a cross-section of the (In,Ga)N/GaN structure. The CL images are recorded at different detection energies representing the emission peaks ``1'' (left) and ``2''  (right CL image) of Fig.\ \ref{fig2}(a). The CL distributions shown in Fig.\ \ref{fig3} confirm illustratively the conclusion drawn from Fig.\ \ref{fig2}, namely that the low- and high-energy emission of the (In,Ga)N originates from the top and bottom part of the (In,Ga)N layer, respectively. The lateral intensity fluctuations of the CL images are partly due to the lateral energy variations shown in Fig.~\ref{fig2}(b). The mean energy values of curves ``1'' and ``2'' of Fig.\ \ref{fig2}(b) represent the average emission energies from the top and bottom part of the layer, which are  $E_t= 2.69$~eV and  $E_b= 3.01$~eV, respectively, in good agreement with the peak positions ``1'' and ``2'' in Fig.\ \ref{fig2}(a).

The average indium content of the top and bottom part of the (In,Ga)N layer can be derived from $E_t$ and $E_b$ using the formula $E_g= 3.42(1-x)+0.67x-bx(1-x)$, where  $E_g$ denotes the band gap energy in eV of (In,Ga)N at room temperature and $b$ the bowing parameter. However, for the bottom part, we have to subtract an energy value $\Delta E_g$, which corrects the blue-shift of the emission energy caused by the compressive strain. The proposed values for this correction range between $\Delta E_g= 1.02x$~eV (McCluskey $et$ $al$.~\cite{McCluskey}) and $\Delta E_g= 1.5x$~eV (Wright $et$ $al$.~\cite{wright}).
 If we use the average indium content of $x= 0.13$ obtained by XRD for the strained part of the layer (cf. Fig.~\ref{fig1}), $\Delta E_g$ ranges between 133 and 195~meV,  which results in a range for the strain-corrected values of $E_b$ between 2.877 and 2.815~eV. For the bowing parameter, we have used a value of $b=1.7$  following the results of Moses and van de Walle.\cite{moses}  With these values, we estimate an indium content between $x=0.128$ and 0.143 for the bottom region of the layer and $x=0.175$ for the region close to the surface.
These values are in good agreement with the ones found by XRD (cf. Fig.~\ref{fig1}). Moreover, the comparison of the indium composition found for the bottom region by CL spectroscopy and XRD shows that the correction of the optical emission with regard to strain results in a better agreement when using the relation proposed by McCluskey $et$ $al$.\cite{McCluskey}

\begin{figure}[!t]
\centerline{\includegraphics*[width=7cm]{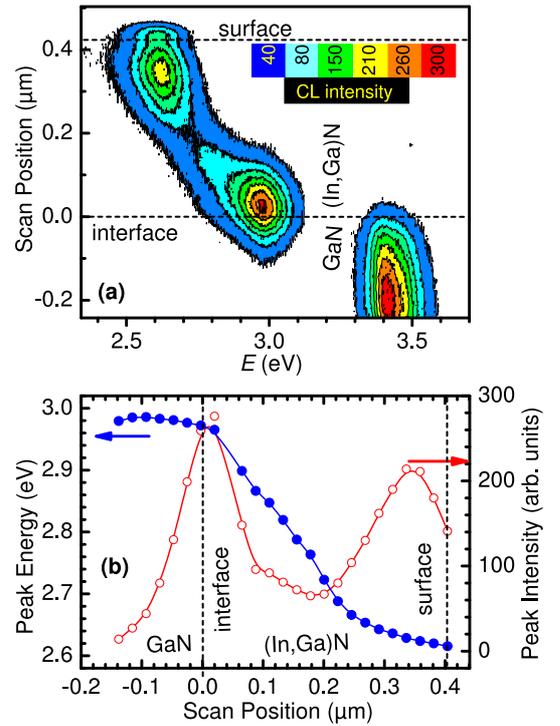}}
\caption{(Color online) (a) CL spectra at 300~K acquired at the cross-section of the (In,Ga)N/GaN structure along a line parallel to the growth direction crossing the interface between the GaN buffer layer and approaching the (In,Ga)N surface. The spectra are presented as a two-dimensional plot with the CL intensity being color coded. (b) Peak energy (dots) and peak intensity (circles) of the (In,Ga)N-related part of the CL spectra shown in (a) as a function of the scan position.} \label{fig4}
\end{figure}

In order to investigate the vertical profile of the luminescence properties and of the indium distribution in detail, we performed spectroscopic CL and EDX measurements on cross-sections of the sample. Figure~\ref{fig4}(a) shows the CL spectra measured along a line (CL line-scan) on the cross-section of the (In,Ga)N/GaN structure starting in the GaN buffer layer and covering the whole (In,Ga)N layer. The spectra are represented as a two-dimensional map, where the CL intensity is color coded. The peak energy and intensity of the (In,Ga)N-related part of these spectra are plotted as a function of the scan position in Fig.~\ref{fig4}(b). Clearly, the energy of the (In,Ga)N-related spectrum decreases gradually from the interface to the surface by about 400~meV. The vertical distribution of the CL intensity exhibits maxima close to the interface as well as near the surface, but it shows a broad minimum centered at a layer depth of about 150~nm. The observation that the (In,Ga)N-related CL is already obtained while the electron beam position is still located within the GaN buffer region (for scan positions $< 0$) is due to the fact that carriers, which have been excited within the GaN layer close to the GaN/(In,Ga)N interface, are able to diffuse to the (In,Ga)N region, where they can recombine radiatively. However, the corresponding CL intensity is assigned to the position of excitation, i.e.,  to a position within the GaN region, reflecting the situation that for the CL measurements only the excitation is local, while the detection integrates over a large volume.
The CL intensity decreases strongly, after the electron beam has reached the (In,Ga)N layer indicating that the relaxation-related evolution of defects already sets in when the layer thickness amounts to values of less than 100~nm. These defects are apparently confined within a region separated from the GaN/(In,Ga)N interface as we have shown previously by TEM investigations.\cite{wang1} A significant part of misfit dislocation segments do not reach the GaN/(In,Ga)N interface. Therefore, the region above the defects is relaxed, whereas the bottom part is still strained.

\begin{figure}
\centerline{\includegraphics*[width=7cm]{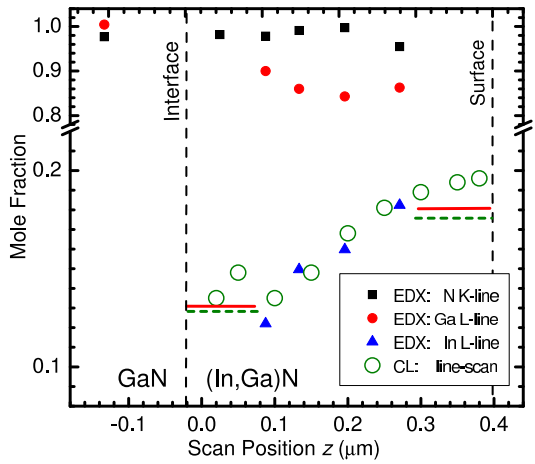}}
\caption{(Color online) Indium content of the (In,Ga)N layer as a function of the distance from the GaN/(In,Ga)N interface. Values represented by circles have been derived from the CL line-scan of Fig.~\ref{fig4}. The other symbols indicate the mole fractions of nitrogen (squares), gallium (dots), and indium (triangles) measured by EDX. The horizontal dashed and solid lines represent the average indium mole fractions determined by CL measurements carried out from the top and by XRD investigations, respectively.} \label{fig5}
\end{figure}

We have selected nine values of the energy profile of Fig.~\ref{fig4}(b) across the whole (In,Ga)N layer and have determined the corresponding indium compositions using the formula and conditions described above. Furthermore, we have corrected the energy for the two positions closest to the GaN/(In,Ga)N interface with regard to the strain according to McCluskey $et$ $al$.\cite{McCluskey} The resulting indium mole fractions are presented by circles in Fig.~\ref{fig5}, where the horizontal solid and dashed lines mark the average $x$ values determined by XRD and CL, respectively. The other symbols represent the mole fractions of nitrogen (K line of N), gallium (L line of Ga), and indium (L line of In) as measured by EDX within the same region as for the CL line-scan. In the vicinity of the surface and interface, mole fractions determined by EDX for $z>0.3$~$\mu$m and  $z<0.1$~$\mu$m are not reliable.
Figure~\ref{fig5} summarizes the experimental results and visualizes the conclusions drawn from our comparative study:

(i) Regarding the equivalence of the exploited methods to determine the indium content, we find a very good agreement between the $x$ values derived from the CL, XRD, and EDX measurements.

(ii) For the part of the layer close to the interface, an appropriate strain correction of the CL emission energy is performed by using the relation $\Delta E_g= 1.02x$~eV proposed by McCluskey $et$ $al$.\cite{McCluskey} Accordingly, the contribution of the strain to the total shift of the CL emission energy along the growth direction of the (In,Ga)N layer amounts to 130~meV.

(iii) The depth profile determined by CL spectroscopy (circles in Fig.~\ref{fig5}) exhibits a variation of $x$ between 0.135 at the bottom and 0.195 at the top of the (In,Ga)N layer.
 Despite the uncertainty in the strain correction within the bottom range, we can state that $x$ remains constant on a level of 0.135 up to a thickness of about 100~nm, increases to a level of 0.19 within a range between 100 and 300~nm, and saturates within the uppermost 100~nm to a level of 0.195. The indium content determined directly by EDX (triangles in Fig.~\ref{fig5}) confirms the gradual increase of $x$ within the central part of the layer very well. Consequently, we conclude that the misfit strain at the growth front decreases gradually until full relaxation is achieved. Subsequently, the indium content does not vary any more as the thickness is further increased.

(iv) The broad minimum of the CL intensity within the scan range of the gradual decrease in emission energy shown in Fig.~\ref{fig4}(b) suggests that a significant amount of relaxation-related defects are created and confined within this region.

(v) The relaxation of the strain does not only result in a pulling of the indium content along the growth direction, but also in a significant enhancement of the lateral variations of $x$ as indicated in Fig.~\ref{fig2}(b) by the much larger energy fluctuation of the CL emission in the upper part of the layer in comparison with the ones from the bottom part of the layer.


\end{document}